\documentstyle[aps]{revtex}
\def\bea{\begin{eqnarray}}
\def\eea{\end{eqnarray}}
\def\nn{\nonumber}
\def\h{\hspace*{.5mm}}
\def\hh{\hspace*{10mm}}
\def\hhh{\hspace*{15mm}}
\def\5{\hspace*{5mm}}
\def\br{{\bf r}}

\def\dr{{\rm d}{\bf r}}

\def\p{\partial}

\def\d{\delta}

\def\la{\langle}
\def\ra{\rangle}
\def\half{\frac12}
\def\x{\xi}

\def\ie{{\it i.e.}}
\begin{document}
\title{Chern--Simons Theory and Quantum Fields in the Lowest Landau Level}
\author{M.\h\/Eliashvili and G.\h\/Tsitsishvili}
\address{Department of Theoretical Physics\\
A.\h\/Razmadze Institute of Mathematics\\
Aleksidze 1, Tbilisi 380093 Georgia\\
{\rm e-mail: simi@rmi.acnet.ge}}
\maketitle
\begin{abstract}
\begin{center}
\parbox{14cm}{
By considering the area preserving geometric transformations in
the configuration space of  electrons moving in the lowest Landau level (LLL)
we arrive at the Chern--Simons type Lagrangian. Imposing the LLL condition,
we get a scheme with the complex gauge fields and transformations. Quantum
theory for the matter field in LLL is considered and formal expressions for
Read's operator and Laughlin wave function are presented in the second
quantized form.}
\end{center}
\end{abstract}

\section{Introduction}

The lowest Landau level comprises the minimal energy states of the charged
particle moving in the $x-y$ plane in the presence of an orthogonal magnetic
field ${\bf B}= {\rm rot}\h{\bf A}$. Corresponding wave functions satisfy
equation
\bea
-iD_{\bar z}({\bf A})\psi_{\rm L}(\br)\equiv
[(\hat p_x+i\hat p_y)-e(A_x+iA_y)]\psi_{\rm L}(\br)=0 \label{eq:lll}
\eea

Electrons in the LLL form a special kind of incompressible quantum fluid
exhibiting such an interesting phenomenon as the fractional quantum Hall
effect (FQHE) (see {\it e.g.} Ref.[1]). A physically attractive way
to understand the formation of corresponding quantum states is the
picture of composite quasiparticles {\it i.e.} fermions (bosons) carrying
odd (even) number of elementary magnetic flux quanta \cite{jain}.
The field-theory realization of this idea leads to the model, where the
matter is interacting with the  Chern--Simons (CS)
gauge field \cite{{lopez},{zhang}}.

It must be noted however that in the most part of presentations
corresponding field theories  include   higher Landau levels
and the subsequent  projection on the LLL is needed \cite{girvin}.
As an interesting alternative one can consider the formulation
including solely  LLL states and fields.

Below we propose a scheme, where the Chern--Simons vector field emerges
as a result of some geometric transformations in the complexified
configuration space of the electron in the LLL. These transformations are
restricted by two conditions: they must be area preserving and the electron
wave function must obey the LLL condition (\ref{eq:lll}). The third assumption
implies that the one-particle charge density distribution does not change
under the about transformation.

The resulted field theory corresponds to the earlier proposed
schemes \cite{{eliashvili},{rajaraman}} with a Chern--Simons field in
the holomorphic gauge  and the field theory realizations of Read's
operator \cite{read} in terms of the non-unitary singular transformation.

In the present paper we reformulate the problem in terms of the fermion
fields obeying the LLL condition (\ref{eq:lll}). We propose the specific
Lagrangian which in the certain limit describes the matter field in the LLL.
As an output we construct the LLL analogue of Read's operator and the
Laughlin wave function \cite{laughlin} for FQHE.

{\bf Notations:}. In the $x-y$ plane together with the Cartesian coordinates
$x^i=x_i$ we use the complex ones: $z=x+iy,\h \bar z=x-iy$ with a metric
tensor $g_{z\bar z}=\half$. Electrons have a charge $e=-1$, mass $m$ and move
in the area $\Omega$ in the homogeneous magnetic field pointing down the
$\hat{\bf z}$ axis: $\p_x A_y-\p_y A_x=-B<0$.

\section{Geometric Transformations}

The classical Lagrangian for the planar electron  moving in
the magnetic field is given by
\bea
L=\frac{m}{2}{\dot x}^i{\dot x}^i-{\dot x}^i\cdot A_i(x).\label{eq:cll1}
\eea
The corresponding canonical Hamiltonian reads
\bea
H_c=\frac{1}{2m}\pi_i\pi_i \equiv\frac{1}{2m}
\left(\pi_z\pi_{\bar z}+\pi_{\bar z}\pi_z\right),\nn
\eea
where $\pi_i=p_i+A_i$ is a kinetic momentum.

In quantum mechanics the wave function of the lowest energy state
satisfies the equation:
\bea
\hat \pi_{\bar z}\psi_{\rm L}({\bf r})\equiv\frac{1}{2}
(\hat\pi_1+i\hat\pi_2)\psi_{\rm L}({\bf r})=0.\label{eq:lll1}
\eea
The solutions to this equation depend on the gauge field ${\bf A}(\br)$.
For example, in the symmetric gauge
\bea
A_1=\frac{B}{2}y,\hh A_2=-\frac{B}{2}x\label{eq:cg},
\eea
which is well adapted to the use of complex variables, the Eq. (\ref{eq:lll1})
looks as follows:
\bea
\left(\frac{\partial}{\partial \bar z}+\frac {B}{4}z\right)
\psi_{\rm L}({\bf r})=0. \nn
\eea
The standard LLL wave function is of the form
\bea
\psi_{\rm L}(z,\bar z)=e^{-(B/4)|z|^2}f(z), \label{eq:wf1}
\eea
with $f(z)$ a function of $z$ only.

The Eq. (\ref{eq:lll1}) can be viewed as a condition imposed on the physical
states by the constraint dynamics (in the  Dirac's \cite{dirac} sense), and at
the classical level these constraints mean the vanishing of the kinetic
momentum:
\bea
\pi_i\approx 0. \label{eq:dc}
\eea

Remark, that one can deduce these constraints starting with the singular
Lagrangian \cite{dunne}
\bea
{\cal L}=-\dot x^i A_i(x), \label{eq:cll2}
\eea
which is a zero mass limit of (\ref{eq:cll1}).

Since the corresponding canonical Hamiltonian vanishes, the quantum dynamics
is completely governed by constraints (\ref{eq:dc}). At the same time one
cannot impose operator constraints
$$
\hat\pi_i|\Psi_{\rm L}\ra=0 \hh \la\Psi_{\rm L}|\hat\pi_i=0,
$$
because they do not commute among themselves
\bea
0=\la\Psi_{\rm L}|[\hat\pi_1,\hat\pi_2]_-|\Psi_{\rm L}\ra=iB\not=0.\nn
\eea
Instead, constraints can vanish only "weakly", {\ie}
\bea
\hat\pi_{\bar z}|\Psi_{\rm L}\ra=0,\hh
\la\Psi_{\rm L}|\hat\pi_z=0. \nn
\eea
This is consistent with classical Eqs. (\ref{eq:dc}), as well as the
corresponding quantum averages vanish
\bea
\la\Psi_{\rm L}|\hat\pi_i|\Psi_{\rm L}\ra=0. \nn
\eea

Consider the zero mass ({\ie} LLL) Lagrangian
\bea
{\cal L}=-\dot x^1(\alpha x^2)-\dot x^2(-\beta x^1) \label{eq:cll3}
\eea
corresponding to the gauge potentials
\bea
A_1=\alpha x^2,\hh A_2=-\beta x^1;\hh\alpha+\beta=B. \nn
\eea

Perform the coordinate transformation
\bea
x^i\rightarrow \nu_{(i)}(x^i+\epsilon^{ik}f_k({\bf r})),\label{eq:tr}
\eea
where ${\bf f}({\bf r})$ is a vector field, and the nonzero $\nu_{(i)}$'s
are scaling factors (there is no summation over the index $(i)$).

Under this transformation the area is changed according to the standard rule
\bea
\Omega \rightarrow \Omega_\nu=\nu\Omega+
\nu \int_{\Omega} \dr \epsilon^{ik}\partial_i f_k({\bf r})
+\nu \int_{\Omega} \dr[\partial_1f_1\cdot\partial_2f_2-
\partial_1f_2\cdot\partial_2f_1] \label{eq:area}
\eea
where $\nu =\nu_{(1)}\cdot \nu_{(2)}$

The Lagrangian (\ref{eq:cll3}) is also transformed
\bea
{\cal L}\rightarrow {\cal L}'_\nu=-\dot x^i(A'_i+a_i)+
\frac{\vartheta}{2}\epsilon^{ik}a_i\dot a_k. \label{eq:lag}
\eea
In the last expression the vector-potential
\bea
A'_1=\nu\alpha x^2,\hh A'_2=-\nu\beta x^1,\hh{\rm rot}\h{\bf A'}(\br)
=\nu B,\nn
\eea
and we have introduced the rescaled vector field
$a_i({\bf r})=-\nu B f_i({\bf r})$ and the constant $\vartheta=(\nu B)^{-1}$.

In the Lagrangian (\ref{eq:lag}) the field ${\bf a}({\bf r},t)$ is treated as
a dynamical variable, hence
$$
\dot {\bf a}=\frac{\rm d}{\rm dt}{\bf a}({\bf r},t).
$$

Now we would like to interpret the transformations (\ref{eq:tr}),
(\ref{eq:area}) and (\ref{eq:lag}) in the light of FQHE characteristics
and identify $\nu$ with the FQHE filling factor
\bea
\nu=\frac{2\pi N}{B\Omega}=\frac{1}{2p+1}.\nn
\eea

Suppose first that the area is not changed, {\ie}
$\Omega\rightarrow\Omega_\nu=\Omega$. This hypothesis is motivated by the
incompressibility of the quantum Hall fluid. Consequently
\bea
(\nu-1)\Omega =\frac{1}{B}
\int{\dr}\epsilon^{ik}\partial_i a_k(\br) \label{eq:area1}
\eea
(here and below we consider transformations (\ref{eq:tr}) with a vanishing
third term on the r.h.s. in Eq. (\ref{eq:area})).

Secondly, consider a particle at ${\bf r}$ in the presence other particles
at ${\bf r}_I$ $(I=1,2,...,N)$, which at the moment, are treated as punctures
in the plane. Associate to the each puncture a singularity, and suppose that
the deformation field $a_i({\bf r})$ satisfies the equation
\bea
\epsilon^{ik}\partial_ia_k({\bf r)}=
\omega\sum_{I=1}^N\delta^{(2)}({\bf r}-{\bf r}_I).\label{eq:rota}
\eea
Eqs. (\ref{eq:area1}) and (\ref{eq:rota}) give
\bea
(\nu-1)\Omega=\frac{\omega}{B}N \nn
\eea
and consequently
\bea
\omega=2\pi\frac{\nu-1}{\nu}=-4\pi p. \nn
\eea
Remark, that the magnetic field $B_1=\nu B $ corresponds to the IQHE with the
filling factor $\nu_1=1$ and suppose, that the wave function corresponding to
the Lagrangian (\ref{eq:lag}) satisfies the LLL condition
\bea
\pi'_{\bar z}|\psi'_\nu\ra=(p_{\bar z}+A'_{\bar z})|\psi'_\nu\ra=0
\label{eq:modc}
\eea
This assumption is equivalent to the following condition imposed
on the field $a_i$:
\bea
a_{\bar z}=0.\label{eq:pol3}
\eea

In the symmetric gauge $(A_z=\frac{i}{4}\bar z,\h A_{\bar z}=-\frac{i}{4} z)$
one easily finds that the sought after vector field ${\bf a}({\bf r})$ can be
presented as a pure gauge
\bea
a_z=\partial_z\Lambda,\hspace{1cm}a_{\bar z}=\partial_{\bar z}\Lambda,\nn
\eea
where
\bea
\Lambda(z|z_I)=
2ip\sum_I \ln(z-z_I).\label{eq:Lambda}
\eea
Indeed, such a vector field guarantees vanishing third term
in Eq. (\ref{eq:area}), has a singular curl (\ref{eq:rota})
and corresponds to Eq. (\ref{eq:pol3}).

It must be emphasized, that the field $a_i({\bf r})$ defined as above,
can not be real. It means that the transformation (\ref{eq:tr}) requires
the complexification of  coordinates. Hence we treat $z$ and $\bar z$ as
independent variables spanning a complex two-dimensional
plane \cite{{nambu},{gaite}} (in the physical subspace $\bar z=z^\star$).

With the help of the singular gauge transformation
\bea
a_i({\bf r})\rightarrow a_i({\bf r})-
\partial_i\Lambda({\bf r}|\br_I) =0 \label{eq:gtr}
\eea
the vector field ${\bf a}$ can be gauged out. In parallel
Lagrangian (\ref{eq:lag}) will be also transformed
\bea
{\cal L}'_\nu\rightarrow {\cal L}_1=-\dot x^iA'_i({\bf r}).  \label{eq:lag1}
\eea
The resulted Lagrangian (\ref{eq:lag1}) corresponds to the LLL electron
in the magnetic field $B_1$.

In quantum mechanics transformations (\ref{eq:gtr}) and (\ref{eq:lag1}) are
accompanied by the wave function transformation
\bea
\psi'_\nu(\br)\rightarrow e^{i\Lambda(\br|\br_I)}\psi'_\nu(\br)
=\psi_1(\br). \nn
\eea
Hence
\bea
\psi'_\nu({\bf r})=\prod_I(z-z_I)^{2p}\psi_1 ({\bf r})\equiv
R(\br|\br_1,..,\br_N)\psi_1 ({\bf r}). \label{eq:rop}
\eea
In the last expression the wave function $\psi'_\nu(\br)$ corresponds to the
Lagrangian (\ref{eq:lag}), $R(\br|\br_1,..,\br_N)$ represents the first
quantized form of Read's operator and $\psi_1(\br)$ is the LLL wave function
of the fermion in the magnetic field $B_1$
\bea
\psi_1(z,\bar z)=e^{-(B_1/4/)z\bar z}f(z). \label{eq:wf2}
\eea

In the space of physical states obeying the LLL condition
(\ref{eq:modc}) operator $\hat\pi'_z$ is given by the expression
\bea
\hat\pi'_z=\hat p_z+A'_z+2ip \sum_I\frac{1}{z-z_I},  \label{eq:pz}
\eea
and with the usual  inner product,
$\hat\pi'_{\bar z}=\hat p_{\bar z}+A'_{\bar z}$
and $\hat\pi'_z$  turn out not to be an
Hermitean conjugate pair of operators.

Now following Ref.[14] modify the definition of the inner product, introducing
\bea
\la\la\Psi|\Psi'\ra=\int {\rm d}^2z\mu(z,\bar z)\psi^\star (z,\bar z)
\psi'(z,\bar z), \nn
\eea
where
\bea
\mu(z,\bar z)=\prod_I|z-z_I|^{-4p}. \label{eq:mes}
\eea
With this inner product $\hat \pi'_{\bar z}$ and $\hat \pi'_ z$ become
Hermitean conjugate to each other
\bea
\la\la\Psi|\hat \pi'_{\bar z}|\Psi'\ra=
\la\la\Psi'|\hat \pi'_ z|\Psi\ra^\star.\nn
\eea

Introduction of the measure (\ref{eq:mes}) can be justified in the another way.
Let us consider the theories corresponding to the Lagrangians (\ref{eq:cll3})
and (\ref{eq:lag}) as equivalent ones. Since under the transformation
(\ref{eq:tr}) the area is preserved, it seems natural to suggest that the
corresponding charge density distribution also remains unchanged. Using
Eqs. (\ref{eq:wf1}), (\ref{eq:rop}) and (\ref{eq:wf2}), for the charge density
distribution one gets the expression
\bea
\varrho({\bf r})=\psi^{\star}_{\rm L}({\bf r})\psi_{\rm L}({\bf r})=
\psi^{\ddag}_\nu({\bf r})\psi_\nu({\bf r}),\nn
\eea
where
\bea
\psi_\nu({\bf r})=
e ^{-(B/4)z\bar z(1-\nu)}\prod_I(z-z_I)^{2p}\psi_1 ({\bf r}),\label{eq:?}
\eea
and
\bea
\psi^{\ddag}_\nu({\bf r})=\mu(z,\bar z)\psi^{\star}_\nu({\bf r})
=\prod_I|z-z_I|^{-4p}\psi^{\star}_\nu({\bf r}). \nn
\eea
With these prescriptions quantum-mechanical averages are defined by
the integral (up to the normalization factor)
\bea
\la\hat {\cal O}\ra=\la\la\Psi_\nu|\hat{\cal O}|\Psi_\nu\ra\equiv
\int {\rm d}^2z\psi^{\ddag}_\nu(z,\bar z)\hat{\cal O}
\psi_\nu(z,\bar z)
\eea
and that constraints (6) vanish in the weak sense
\bea
\la\la\Psi_\nu|\hat\pi_i|\Psi_\nu\ra=0 .\nn
\eea

Now one can ask the question: what is the  field theory counterpart of the
above construction? From the Lagrangian (\ref{eq:lag}) it follows the
Euler--Lagrange equation for the field $a_i$:
\bea
\dot a_k=-\frac {1}{\theta}\epsilon_{kl}\dot x^l.
\eea
Substituting this  in  (\ref{eq:lag}) we arrive at the effective Lagrangian
\bea
{\cal L}_{eff}=-\dot x^i(A'_i+\half a_i), \label{eq:leff}
\eea
where the curl of  $a_i({\bf r})$ satisfies (\ref{eq:rota}).

It is not difficult to guess, that the corresponding field theory is
given by the CS gauge Lagrangian:
\bea
{\cal L}_{CS}=-j^\mu(x)(A'_\mu(x)+a_\mu(x))+
\frac{\kappa}{2}\varepsilon^{\mu\nu\lambda}a_\mu(x)\partial_\nu a_\lambda(x)
\label{eq:lcs}
\eea
(the Greek indices run over values 0, 1, 2).

This equivalence can be  demonstrated easily, plugging the field equations
\bea
\kappa \varepsilon^{\mu\nu\lambda}\partial_{\nu} a_\lambda(x)=
j^\mu(x) \nn
\eea
back into (\ref{eq:lcs}). In the temporal gauge $a_0=0$
the corresponding effective Lagrangian
\bea
{\cal L}_{eff}=-j^\mu(x)(A'_\mu(x)+\half a_\mu(x)) \label{eq:leff1}
\eea
is an exact analogue of (\ref{eq:leff}). The Gauss law
\bea
\epsilon^{ik}\partial_i a_k(x)=\frac{1}{\kappa}j^0(x)
\eea
is identical with (\ref{eq:rota}) if one puts:
\bea
\kappa=-\frac{1}{4\pi p}=-\frac{N}{2p\Omega}\theta.  \label{eq:41}
\eea

So one can reconstruct the gauge Lagrangian with the Chern--Simons constant,
determined by the requirement of the area conservation under the coordinate
transformations (\ref{eq:tr}).

\section{Read's Operator and Field Theory in LLL}

In this item we would like to obtain the second quantized counterpart of
the Eq. (\ref{eq:rop}). The problem is that the space of LLL states is not
complete and there is no direct way to write down the particle density
operator with $\delta$-type singularities corresponding to point-like
excitations.

Denote by $\hat\psi_\nu(\br)$ and $\hat\psi_1(\br)$ the quantum fields
corresponding to the wave functions under consideration. The matter fields
in the lowest Landau level can be expanded in the series
\bea
\hat\psi_1(\br)=\sum_n^{\infty} \hat f_nu_n(\br)\hhh
\hat\psi_1^\dagger(\br)=\sum_n^{\infty} \hat f^+_nu^\star _n(\br),\nn
\eea
where the system of orthonormal solutions $ u_n ({\bf r})$ in the symmetric
gauge is given by
\bea
u_n(\br)=\frac{1}{\sqrt{2\pi}\ell}\frac{1}{\sqrt{2^nn!}}
\left(\frac{z}{\ell_1}\right)^ne^{-(1/4)|z/\ell_1|^2},\hhh
\ell_1^2=\frac{1}{B_1}, \nn
\eea
and  Fermi amplitudes satisfy standard relations
\bea
\hat f_n\hat f^+_m+\hat f^+_m\hat f_n=\delta_{nm}.\nn
\eea

The anti-commutator for Fermi fields in the LLL
is expressed in terms  of the bilocal kernel
\bea
[\hat\psi_1(\br'),\hat\psi_1^\dagger(\br'')]_+=
\sum_n^\infty u_n(\br' ) u^\star_n(\br'')=K_1(\br'|\br'') \label{eq:ker}
\eea
instead of the customary $\delta({\bf r}'-{\bf r}'')$ function. Hence the
conventional charge density
$\hat\varrho(\br)= \hat\psi_1^\dagger(\br)\hat\psi_1(\br)$
is not a good candidate for the sought after  density operator.

Being the projection onto the space of LLL states, the operator
$K_1(\br'|\br'')$ is not invertible. One can bypass this obstacle
introducing the modified kernel
\bea
K_\x(\br'|\br'')=\x K_1(\br'|\br'')+
(1-\x)[\d(\br'|\br'')-K_1(\br'|\br'')],
\label{eq:mk}
\eea
which coincides with (\ref{eq:ker}) in the limit $\x\to 1$.
For $\x\not=1$ exists an inverse operator:
\bea
K^{-1}_\x(\br'|\br'')=\frac{1}{\x}K_1(\br'|\br'')+\frac{1}{1-\x}
[\d(\br'|\br'')-K_1(\br'|\br'')]. \label{eq:invk}
\eea
Now consider the action
\bea
{\cal A}_\x=i\int{\rm dt}\dr'\dr''\psi_\x^\star(t,\br')
K^{-1}_\x(\br'|\br'')
\p_t\psi_\x(t,\br'')-
\int{\rm dt}\dr \psi_\x^\star(t,\br)V(\br)\psi_\x(t,\br) \label{eq:ma}
\eea
where $V(\br)$ is some one-particle potential.

Varying this action with respect to $\psi_\x^\star(t,\br)$ one gets
the $\x$-dependent  equation
\bea
i\int\dr'K^{-1}_\x(\br|\br')\p_t\psi_\x(t,\br')
= V(\br)\psi_\x(t,\br),  \hh \x\not= 1.
\eea
The last equation can be inverted
\bea
i\p_t \psi_\x(t,\br)=
\int\dr' K_\x (\br|\br') V(\br')\psi_\x(t,\br'). \label{eq:el}
\eea
Thus we arrive at the Euler--Lagrange equation of motion for
the field $\psi_\x(t,\br)$ corresponding to the action (\ref{eq:ma}).

In the last expression one can readily go to the limit $\x \to 1$ and obtain
the single-particle Schr$\ddot{\rm o}$dinger equation for the matter field in
the lowest Landau level
\bea
i\partial_t \psi_1(t, {\bf r})=
\int {\rm d}^2{\bf r'}K_1({\bf r}, {\bf r}')V({\bf r'})\psi_1( t,{\bf r'}).
\label{eq:stone2}
\eea
This equation first was derived by J. Martinez and M. Stone \cite{martinez}
using the variational formalism for the constrained system with the Lagrangian
\bea
L=\int {\rm d}^2{\bf r}\psi_1^\star( t,{\bf r})(i\partial_t-V({\bf r}))
\psi_1(t, {\bf r})\nn
\eea
\bea
+\int {\rm d}^2{\bf r}([\psi_1(t, {\bf r})-
\int {\rm d}^2{\bf r}'K_1({\bf r}, {\bf r}')\psi_1(t, {\bf r}')
\eta(t,{\bf r}')]+h.c). \nn
\eea
Here $\eta({\bf r})$ is
the Lagrange multiplier introduced in order to enforce the LLL condition
\bea
\psi_1(t,\br)=\int \dr'  K_1(\br|\br')\psi_1(t,\br') \label{eq:ms}
\eea
(the last relation is equivalent to $D_{\bar z}({\bf A'})\psi_1(\br)=0$).

The canonical momentum is given by the nonlocal expression
\bea
\pi_\x(t,\br)=\frac{\delta L}{\delta(\p_t\psi_\x(t,\br))}=
i\int\dr'\psi_\x^*(t,\br')K^{-1}_\x(\br'|\br). \nn
\eea
 From the equal time anti-commutators
\bea
[\hat\pi_\x(t,\br),\hat\psi_\x(t,\br')]_+=i\delta(\br-\br'), \hh
[\hat\psi_\x(\br),\hat\psi_\x^\dagger(\br')]_+=M_\x(\br|\br') \nn
\eea
one  sees, that the operator
\bea
\hat R_\x(t,\br)=-i\hat\pi_\x(t,\br)\hat\psi_\x(t,\br)\equiv
\int\dr'\hat\psi_\x^\dagger(t,\br')K^{-1}_\x(\br'|\br)\hat\psi_\x(t,\br)
\label{eq:hr}
\eea
has a $\delta$-like commutator with the matter field:
\bea
[\hat R_\x(t,\br),\hat\psi_\x(t,\br')]_-=
-\delta(\br-\br')\hat\psi_\x(t,\br). \label{eq:rc}
\eea

Now one can suggest the limiting  procedure, implying that
$\lim_{\x\to 1}\hat\psi_\x(\br)=\hat\psi_1(\br)$ is a LLL operator.
Below we will operate with the needed quantities and relations for
$\x\not= 1$, and go to the limit $\x\rightarrow 1$ only in the final
expressions. Note that it is a usual procedure performed with the gauge
fixing term in gauge theories. From the functional integral point of view
the main contribution comes from the configurations satisfying the LLL
condition (\ref {eq:ms}), as well as $\frac{1}{1-\x}$ terms are rapidly
oscillating.

For $\x\not=1$ one can define the field operator
\bea
\hat\psi_{\nu,\x}(\br)=e^{\Phi_\x(\br)}\hat\psi_\x(\br)\nn
\eea
where
\bea
\Phi_\x(\br)=
-2p\frac{B_1}{4}|z|^2 +
2p\int \dr'\ln (z-z')\hat R_\x(\br'), \nn
\eea
and fields are taken at the fixed time (say $t=0$).

Define the "vacuum" state $\la0_\x|\hat\psi_\x^\dagger(\br)=0$ and consider
the bra-vector:
\bea
\la0_\x|\hat\psi_\x(\br_1)\cdots\hat\psi_\x(\br_N)\cdot\hat\psi_{\nu,\x}(\br)=\nn
\eea
\bea
=\la0_\x|e^{ 2p\int \dr'\ln (z-z')\hat R_\x(\br')}
e^{i\Lambda(z|z_I)}
\hat\psi_\x(\br_1)\cdots\hat\psi_\x(\br_N)\cdot\hat\psi_\x(\br)=\nn
\eea
\bea
=e^{i\Lambda(z|z_I)}
\la0_\x|\hat\psi_\x(\br_1)\cdots\hat\psi_\x(\br_N)\cdot\hat\psi_\x(\br).
\eea
In the last expression one can safely go to the limit $\x=1$ recovering
the second quantized analogue of the Eq. (\ref{eq:rop}).

At the same way one can consider the state:
\bea
\la0_\x|\hat\psi_{\nu,\x}(\br_1)\cdots\hat\psi_{\nu,\x}(\br_N)=
e^{-2p(B_1/4)\sum |z_I|^2}\prod_{I<K}(z_I-z_K)^{2p}
\la0_\x|\hat\psi_\x(\br_1)\cdots\hat\psi_\x(\br_N). \nn
\eea
Introduce the ground state corresponding to fermions in the magnetic
field $B_1$, totally filling up lowest Landau level
\bea
|GS\ra =\prod_{j=0}^{N-1}f^+_j|0\ra,\hhh N=\frac{B_1\Omega}{2\pi}. \nn
\eea
Then
\bea
\lim_{\x \to 1}\la 0_\x|\hat\psi_\x(\br_1)\cdots\hat\psi_\x(\br_N)|GS\ra=
e^{-(B_1/4)\sum |z_I|^2}\prod_{I<K}(z_I-z_K)
\eea
is the corresponding $\nu=1$ wave function (Slater determinant).

Consequently, the limiting value of the matrix element
\bea
\lim_{\x \to 1}\la 0_\x|\hat\psi_{\nu,\x}(\br_1)\cdots\hat\psi_{\nu,\x}
(\br_N)|GS\ra=
e^{-(B/4)\sum |z_I|^2}\prod_{I<K}(z_I-z_K)^{2p+1} \label{eq:LWF}
\eea
reproduces the Laughlin wave function \cite{laughlin} for the fractional
states with $\nu=\frac{1}{2p+1}$.

\section{Concluding Remarks}

The aim of the present paper is two-fold. First, we tried to reveal the link
between area preserving diffeomorphisms in FQHE \cite{{capelli},{karabali}}
and CS gauge fields. Demanding that these transformations does not violate
the quantum-mechanical LLL condition we arrive to the conclusion, that the
CS field corresponds to the non-unitary similarity transformation between the
integral and fractional quantum Hall systems \cite{{eliashvili},{rajaraman}}.

Secondly, we attempted to find a field-theory form for the singular gauge
transformations in terms of purely LLL field operators. Although we failed
to give a direct solution to this task, an extended Lagrangian formalism and
the limiting procedure is described, permitting to reproduce the needed second
quantized quantities. Remark that the proposed scheme agrees with the earlier
one \cite{martinez} in the general aspects.

Concluding, we would like to note, that the field-theory constructions in
the LLL (like Read's operator, Jains composite particles) can be naturally
incorporated in the CS gauge theory with the complex gauge transformation
group. In the present paper we discuss a version of this scheme, where the
complex gauge transformations are induced by  area preserving coordinate
transformations in the lowest Landau level. It seems interesting to consider
the same framework for bilayer quantum Hall systems and point particles with
non-Abelian charges.

\section{Acknowledgments}

We thank G. Japaridze for many useful discussions. M.E. is very grateful to
P. Sorba for his hospitality at LAPTH (Annecy), where the part of the present
work was done. Work was supported in part by the grant INTAS-Georgia 97-1340
and by Georgian Academy of Sciences, under grant No. 1.4

\end{document}